\documentclass[prl,aps,twocolumn,showpacs,amssymb]{revtex4}
\usepackage{color,graphicx,shortvrb}
\usepackage{amsfonts}
\usepackage{tabularx}
\usepackage{dcolumn}

\begin{document}
\title{Very high precision bound state spectroscopy near a
$^{85}$Rb Feshbach resonance} \author{N.~R. Claussen,
S.~J.~J.~M.~F.~Kokkelmans$^*$, S.~T. Thompson, E.~A. Donley, and
C.~E. Wieman} \affiliation{JILA, National Institute of Standards
and Technology and the University of Colorado, and the Department
of Physics, University of Colorado, Boulder, Colorado 80309-0440}

\date{\today}

\begin{abstract}
We precisely measured the binding energy ($\epsilon_{\rm bind}$)
of a molecular state near the Feshbach resonance in a $^{85}$Rb
Bose-Einstein condensate (BEC). Rapid magnetic field pulses
induced coherent atom-molecule oscillations in the BEC.  We
measured the oscillation frequency as a function of B-field and
fit the data to a coupled-channels model.  Our analysis
constrained the Feshbach resonance position [155.041(18)~G], width
[10.71(2)~G], and background scattering length [-443(3)~a$_0$] and
yielded new values for $v_{DS}$, $v_{DT}$, and $C_6$. These
results improved our estimate for the stability condition of an
attractive BEC. We also found evidence for a mean-field shift to
$\epsilon_{\rm bind}$.
\end{abstract}


\maketitle The phenomenon of a Feshbach resonance in ultracold
collisions of alkali atoms has received much theoretical and
experimental interest in recent years and has sparked interest in
the subject of resonant Bose-Einstein condensates (BEC). Feshbach
resonances \cite{feshbach1958a,Tiesinga1992a} have been used to
control elastic and inelastic collisions in ultracold gases
\cite{Courteille1998a,Roberts1998a,Vuletic1999a,Loftus2002a} and
for tuning the self-interaction in BEC
\cite{Inouye1998a,Cornish2000a,Donley2001a,Khaykovich2002a,Strecker2002a}
by changing the magnitude of an external magnetic field.  The
magnetic field controls the self-interaction in the BEC by
affecting the s-wave scattering length, $a$.  Close to resonance,
the scattering length varies with B-field according to
\begin{equation} a=a_{\rm bg} \left( 1-\frac{\Delta}{B-B_{\rm
peak}}\right), \label{equ1} \end{equation} where B$_{\rm peak}$ is
the resonance position and is defined to be the magnetic field
where the magnitude of $a$ becomes infinite, $a_{\rm bg}$ is the
background scattering length, and $\Delta=B_{\rm zero}-B_{\rm
peak}$ is the resonance width where $B_{\rm zero}$ is the B-field
where the scattering length crosses zero. Measurements of Feshbach
resonance positions and widths have been used in a variety of
alkali atoms to improve the determination of the interatomic
potentials.  These potentials have then been used to precisely
calculate a multitude of important properties for trapped atomic
gases \cite{Roberts1998a,
Loftus2002a,vanAbeelen1999b,Leo2000a,Roberts2001b,vanKempen2002a}.

Recently, we applied rapid magnetic field variations near a
Feshbach resonance to create an atom-molecule superposition state
in a $^{85}$Rb BEC \cite{Donley2002a}, which has allowed us to
precisely determine the Feshbach resonance position and width. Our
novel technique for studying the Feshbach resonance relies on the
presence of atom-molecule coherence
\cite{Kokkelmans2002a,Kohler2002b,Mackie2002b}.  By inducing
periodic oscillations in the number of condensate atoms, we obtain
a direct, high precision measurement of the molecular bound state
energy. Exploiting the resonance, we tune the molecular state very
close to threshold --- to our knowledge, this is the most weakly
bound state ever observed. The present method for studying the
Feshbach resonance through atom-molecule oscillations offers all
of the many inherent advantages of a frequency measurement,
including the possibility of high measurement precision, a lack of
sensitivity to errors in the absolute atom number calibration, and
a simple interpretation of the oscillation frequency in terms of
the relative energy difference between the atomic and molecular
states.  When these advantages are combined with an improved
method for magnetic field calibration \cite{Claussen2002a}, the
present technique for probing the Feshbach resonance is much more
precise than previous experiments that examined such Feshbach
resonance observables as variable rethermalization rates in a
trapped cloud of atoms \cite{Roberts1998a,Roberts2001b},
enhancements of photoassociation rates \cite{Courteille1998a} and
inelastic loss rates near the resonance \cite{Chin2000a}, and
variations of the mean-field expansion energy of a BEC
\cite{Inouye1998a}.

To complete our precise characterization of the Feshbach
resonance, we also made an improved measurement of B$_{\rm zero}$,
the magnetic field where the scattering length vanishes.  This
experiment is very similar to our previous work
\cite{Roberts2001a,Roberts2001b}, where we determined the $a$=0
field by measuring the critical number (N$_{\rm crit}$) for
collapse of a BEC, and then we extrapolated to the magnetic field
where N$_{\rm crit}$ would be infinite.  We have improved the
measurement precision by about a factor of 4 by improving our
magnetic field calibration and using a larger number of condensate
atoms to measure the collapse. We find B$_{\rm
zero}$=165.750(13)~G.

The procedure used to generate atom-molecule oscillations in
$^{85}$Rb Bose-Einstein condensates has been described in previous
work \cite{Donley2002a}, so we merely outline the method here.
After creating condensates with initial number of atoms N$_0
\simeq$16000 at a magnetic field B$\simeq$162~G, we apply two
short B-field pulses ($\sim$40~$\mu$s duration) that approach and
then recede from the Feshbach resonance at B$_{\rm peak}
\simeq$155~G. The intermediate value of magnetic field between the
pulses, B$_{\rm evolve}$, and the time spacing between pulses,
t$_{\rm evolve}$, are variable quantities. The double pulse
sequence is followed by a slow change in the B-field to expand the
BEC \cite{Donley2001a}, then the trap is switched off
(B$\rightarrow$0) and destructive absorption imaging is used to
count the number of atoms remaining in the condensate.

As in Ref.~\cite{Donley2002a}, periodic oscillations in the BEC
number were observed as a function of t$_{\rm evolve}$ (see
Fig.~1).  We fit the BEC number oscillation to a damped harmonic
oscillator function with an additional linear loss term:
\begin{equation} N(t) = N_{avg} - \alpha t + A \exp{(-\beta t)}
\sin{(\omega_{e} t + \phi)}, \end{equation} where $N_{avg}$ is the
average number, A is the oscillation amplitude, $\alpha$ and
$\beta$ are the number loss and damping rates, respectively, and
$\omega_{e}=2 \pi \sqrt{\nu_0^2-[\beta/(2 \pi)]^2}$.  The quantity
of interest here is $\nu_0$, the natural oscillator frequency
corresponding to the molecular binding energy,
$\nu_0$=$\epsilon_{\rm bind}/h$. We measured the oscillation
frequency for values of B$_{\rm evolve}$ from 156.1~G to 161.8~G.
Over this range, the frequency varies by over 2 orders of
magnitude (10-1000 kHz), but the linear loss rate changes very
little.  The damping rate shows a significant B-field dependence,
increasing from $\beta \simeq$ 2$\pi$~x~0.8~kHz near 156~G to
$\beta \simeq$ 2$\pi$~x~22~kHz near 162~G.  We find no significant
density dependence to the damping at B$_{\rm evolve}$=158.60(5)~G;
increasing the total atom density by a factor of 4.3(5) leads to a
damping rate increase of only a factor of 1.3(3).  Atom loss from
the BEC is well described by a linear rate of -2 to
-7~atoms/$\mu$s over the field range of interest. The rate is
consistent with previous measurements of number loss due to a
single B-field pulse toward the Feshbach resonance
\cite{Claussen2002a}.

\begin{figure} \includegraphics[bb=30 10 555 730,clip,scale=0.45]{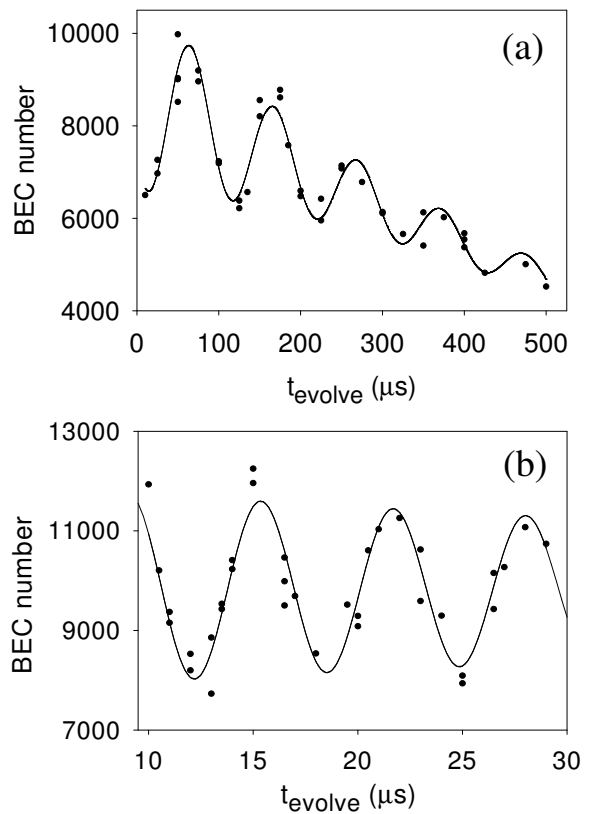}
\caption{BEC number versus pulse spacing, t$_{\rm evolve}$.
{\textbf (a)} B$_{\rm evolve}$=156.840(25)~G. At this magnetic
field, which is relatively close to resonance, the oscillation
frequency is very low ($\nu_0$=9.77(12)~kHz) so that the damping
and atom loss significantly affect the observed time dependence
(here $\beta$=2$\pi \times 0.58(12)$~kHz and
$\alpha$=7.9(4)~atoms/$\mu$s). {\textbf (b)} B$_{\rm
evolve}$=159.527(19)~G. Farther from resonance, the time
dependence of the BEC number is dominated by the higher
oscillation frequency of $\nu_0$=157.8(17)~kHz. Damping of the
oscillations and atom loss are negligible in the relatively short
time window used to determine $\nu_0$.} \label{fig-oscillate}
\end{figure}

To characterize the Feshbach resonance, it is necessary to know
both the oscillation frequency and B$_{\rm evolve}$. We precisely
measured B$_{\rm evolve}$ by transferring atoms to an untrapped
spin state by driving $\Delta$m=+1 spin flip transitions with an
applied pulse of rf radiation (pulse
length=5$\rightarrow$25~$\mu$s).  The spin flip frequency was
determined from the rf lineshape for the loss of atoms from the
magnetic trap.  After measuring the rf transition frequency, we
inverted the Breit-Rabi equation to obtain the corresponding
B-field.  To ensure that the magnetic field was sufficiently
constant during t$_{\rm evolve}$, we mapped out $B(t)$ using rf
pulses with lengths short compared to t$_{\rm evolve}$. Due to
interference of the rf radiation with the magnetic field control
circuitry, there was a small systematic shift of the field as a
function of the rf power used. The uncertainty for each magnetic
field determination was the quadrature sum of the uncertainty due
to the lineshape measurements ($\sim$15~mG) and the uncertainty in
the extrapolation to zero rf power ($\sim$20~mG). Typically, the
total uncertainty in the average B-field was $\sim$25~mG.

The measured oscillation frequencies versus magnetic field are
plotted in Fig.~2. We use these data and the zero-crossing field
$B_{\rm zero}$ to completely characterize the scattering length
and binding energy as a function of magnetic field near the
Feshbach resonance.  As a starting point we use the
coupled-channels analysis of
van~Kempen~et~al.~\cite{vanKempen2002a}, where several
high-precision data for $^{85}$Rb and $^{87}$Rb were combined to
perform an inter-isotope determination of the rubidium
interactions with unprecedented accuracy. The predictive power of
this analysis can be seen from Ref.~\cite{Donley2002a}, where the
initial data on the atom-molecule coherence were already in good
agreement with the predicted binding energy of the underlying
Feshbach state. Another example of the accuracy of the analysis in
Ref.~\cite{vanKempen2002a} is its agreement with more than 40
Feshbach resonances recently discovered in $^{87}$Rb
\cite{marte2002a}.

\begin{figure}
\includegraphics[bb=30 10 555 730, clip,scale=0.45]{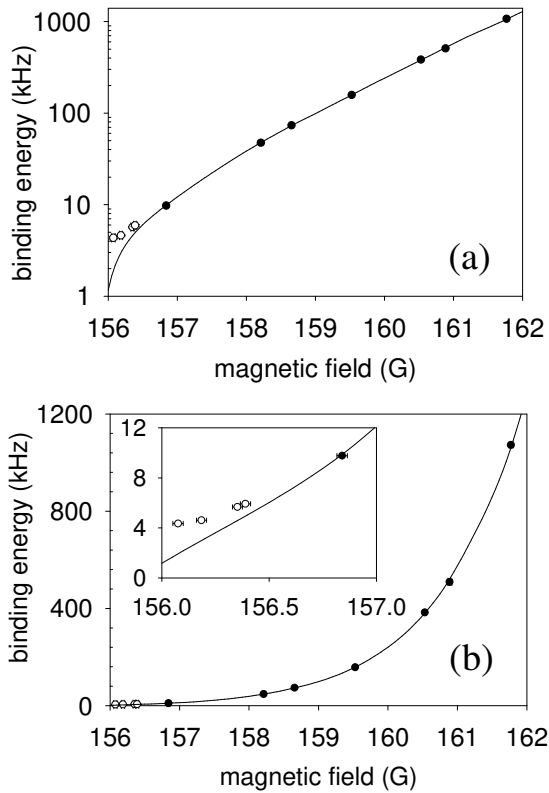}
\caption{Molecular binding energy versus magnetic field, B$_{\rm
evolve}$. {\textbf (a)} The points are measured values of the
atom-molecule oscillation frequency, $\nu_0$, while the solid line
represents the molecular binding energy, which we fit to the data
by adjusting the parameters of a coupled-channels scattering
theory. Only black points were included in the fit; white points
were excluded because they experienced a statistically significant
mean-field shift.  To improve visibility, the points are larger
than the error bars. {\textbf (b)} Same as in {\textbf (a)}, but
with a linear scale for the vertical axis.  The inset shows the
deviation of the lowest frequency data from the fit to the rest of
the data. } \label{fig-ebind} \end{figure}

Van~Kempen~et~al. used the best known values \cite{Roberts2001b}
for the resonant magnetic field $B_{\rm peak}$ and zero crossing
$B_{\rm zero}$.  In this Letter, we ignore the relatively
imprecise value of B$_{\rm peak}$ from Ref.~\cite{Roberts2001b},
and instead use the measured dependence of binding energy on
magnetic field along with the new B$_{\rm zero}$ measurement given
above to determine the interaction parameters. We observe that the
fitting procedure is mainly sensitive to only three parameters:
the van der Waals dispersion coefficient, $C_6$, and the
non-integral vibrational quantum numbers at dissociation, $v_{DS}$
and $v_{DT}$, which determine the position of the last bound state
in the singlet and triplet potentials, respectively. Varying the
additional parameters $C_8$, $C_{10}$, $\phi_T^E$ (the first-order
energy-dependence of the phase of the oscillating triplet radial
wave function), and $J$, the strength of the exchange interaction,
does not improve the fitting because these changes can be absorbed
in small shifts of $v_{DS}$, $v_{DT}$, and $C_6$. Therefore, we
take the mean values for these four parameters~\cite{explain-phi}
from the most recent determination in Ref.~\cite{marte2002a}.

The best fit to B$_{\rm zero}$ and the seven highest frequency
data points yields a reduced $\chi^2$=0.30 for 5 degrees of
freedom. This value of $\chi^2$ is improbably low due to the fact
that the uncertainty in the data is dominated by the systematic
uncertainty in magnetic field related to the magnitude of the rf
power shift. Figure~2 shows the theoretical fit to the binding
energy data as a function of magnetic field. From the fit, we find
substantially improved values for the Feshbach resonance position
$B_{\rm peak}$=155.041(18)~G, width $\Delta$=10.71(2)~G, and
background scattering length $a_{\rm bg}$=-443(3)~a$_0$. These
results may be compared to previously obtained results $B_{\rm
peak}$=154.9(4)~G and $\Delta$=11.0(4)~G~\cite{Roberts2001b}, and
$a_{\rm bg}$=-450(3)~a$_0$ \cite{explain-abg}. Our best
interaction parameter values are $C_6$=4707(2)~a.u.,
$v_{DS}$=$0.00918(17)$, and $v_{DT}$=$0.94659(29)$.  Here the
error bars do not include systematic errors due to the
uncertainties in other interaction parameters that are not
constrained by our data. To compare our values with those of
Ref.~\cite{vanKempen2002a}, we determined the sensitivity of our
three interaction parameters to systematic shifts in the other
parameters, as shown in Table~I. Using the fractional
uncertainties in $C_8$, $C_{10}$, $\phi_T^E$, and $J$ from
Ref.~\cite{vanKempen2002a}, we find $C_6$=4707(13)~a.u.,
$v_{DS}$=$0.0092(4)$, and $v_{DT}$=$0.9466(5)$.  All of these
values agree with those given in Ref.~\cite{vanKempen2002a}:
$C_6$=4703(9)~a.u., $v_{DS}$=$0.009(1)$, and $v_{DT}$=$0.9471(2)$.
Our value for $v_{DS}$ is more precise than that of
Ref.~\cite{vanKempen2002a}, while $v_{DT}$ and $C_6$ are slightly
less precise.  If future experiments allow improvements to the
other interaction parameters, then our results will also become
more precise since the systematic errors are comparable to or
larger than our statistical errors from the fit.

\begin{table}[t!]
\caption{Sensitivities of the determined interaction parameters
$v_{DS}$, $v_{DT}$ and $C_6$ to fractional uncertainties in $C_8$,
$C_{10}$, $\phi_T^E$ and $J$.  For instance, the systematic error
in $C_6$ due to a 10$\%$ uncertainty in $C_8$ is $123 \times 0.10
= 12.3$~ a.u.}
\begin{ruledtabular}
\begin{tabular}{ccccc}
& $\Delta C_8/C_8$ & $\Delta C_{10}/C_{10}$ & $\Delta
\phi_T^E/\phi_T^E$ &   $\Delta J/J$ \\

\hline $\Delta v_{DS}$ & $-1.53 \times 10^{-4}$ & $-6.80\times
10^{-5}$ & $-2.59\times 10^{-3}$ &   $1.72\times 10^{-3}$ \\

$\Delta v_{DT}$ & $-4.14\times 10^{-4}$ & $-1.39\times 10^{-4}$ &
$2.31\times 10^{-3}$ &   $1.71\times 10^{-3}$\\

$\Delta C_6$ & $123$ & $33.4$ & $-47.8$ & $19.3$ \\
\end{tabular}
\end{ruledtabular}
\label{tab1}
\end{table}

To understand the strong parameter constraints that we obtain with
our bound state spectroscopy, it is important to consider the
nonlinear dependence of the binding energy on magnetic field. The
magnetic field dependence of $\epsilon_{\rm bind}$ as it
approaches the collision threshold depends sensitively on the
exact shape of the long range interatomic potentials, which are
mainly characterized by the van der Waals coefficient, $C_6$.  At
magnetic fields far from resonance, the bound state wave function
is confined to short internuclear distance and the binding energy
varies linearly with magnetic field.  The linear dependence on
B-field gives relatively little information about $C_6$.  As the
B-field approaches resonance, the detuning decreases until the
bound state lies just below threshold.  Now the bound state wave
function penetrates much deeper into the classically forbidden
region, which causes $\epsilon_{\rm bind}$ to curve toward
threshold as a function of magnetic field.  Because the
energetically forbidden region stretches out as $C_6/r^6$, the
observed curvature depends sensitively on the $C_6$ coefficient.
One can show~\cite{kokkelmans2002tb} that an analytical Feshbach
model that includes the correct potential range and background
scattering processes~\cite{Kokkelmans2002b} can reproduce the
binding energy curve over the full range of magnetic field.

The coupled-channels theory used in this work applies to two-body
scattering; therefore, this theory cannot account for many-body
effects in the atom-molecule BEC system, such as a mean-field
shift to the observed oscillation frequency
\cite{Kokkelmans2002a,Duine2002a}. Any such mean-field shift must
be fractionally largest near the Feshbach resonance, where the
binding energy approaches zero while the atom-atom scattering
length increases to infinity. We searched for a mean-field shift
to the oscillation frequency when B$_{\rm evolve}$ was decreased
to $\sim$156~G.  As shown in Fig.~2, the lowest magnetic field
data display a clear frequency shift with respect to the
coupled-channels theory prediction.  As B$_{\rm evolve}$
approaches resonance, the observed shift increases to 1.7~kHz,
which significantly exceeds a simple estimate for the average
atom-atom mean-field shift in the BEC: $4 \pi \hbar^2 \langle n
\rangle a/m \simeq 0.5$~kHz at B$_{\rm evolve}$=156.1~G.  We are
presently investigating new experimental techniques to further
study the mean-field shifts, including their density dependence.

Since the lowest frequency data show evidence for a mean-field
shift, we exclude these points from the (two-body) theory fit.  We
determine the cutoff magnetic field for the excluded region by the
following procedure.  We fit the data set that includes all
frequency measurements satisfying $\nu_0 \geq 9$~kHz.  Eliminating
the lowest frequency point from the set causes the reduced
$\chi^2$ to decrease from 0.3 to 0.2, and there is no significant
change in parameter values.  In contrast, adding the next lower
frequency point increases the reduced $\chi^2$ to 1.9, causing a
systematic shift in the parameter values.  The observed behavior
seems sensible since we expect mean-field shifts to increase
rapidly as one moves toward resonance (see Fig.~2).

As a result of the improved determination of the $^{85}$Rb
Feshbach resonance parameters, we find that our new value for the
off-resonant or background scattering length, $a_{\rm
bg}$=-443(3)~a$_0$, is inconsistent with the value given in
Ref.~\cite{Roberts2001b}, where $a_{\rm bg}$=-380(21)~a$_0$.  The
most plausible explanation we can find for disagreement is that
the theoretical expression used to relate measured
rethermalization rate to cross section is insufficient for the
requisite level of accuracy.  However, the new value for $a_{\rm
bg}$ allows us to revise our previous estimate for the stability
condition of a BEC with negative scattering
length~\cite{Roberts2001a}.  We use Eq.~(\ref{equ1}) to obtain the
linear slope of scattering length versus B-field near B=B$_{\rm
zero}$. We then find the stability coefficient for BEC collapse,
$k_{\rm collapse}$, by combining the value of $\Delta a/\Delta
B$=-39.87(22)~a$_0$/G with the measured slope of 1/N$_{\rm crit}$
versus magnetic field~\cite{Roberts2001a} of 0.00126(3) (atoms
G)$^{-1}$. Thus, we obtain the revised value $k_{\rm
collapse}$=0.547(58), where the error is dominated by a 10\%
systematic uncertainty in the determination of N$_{\rm crit}$. The
present determination agrees with the theoretical value of
0.55~\cite{Gammal2001a}.

In conclusion, we present a unique method for exploring a
$^{85}$Rb Feshbach resonance.  The observed atom-molecule
coherence allows us to study the highly nonlinear dependence of
the molecular binding energy on magnetic field.  We find good
agreement with an analysis of van~Kempen~et al.
\cite{vanKempen2002a} and improved precision for the
characterization of the Feshbach resonance. In addition, we
observe mean-field shifts to the binding energy, offering the
possibility for future studies of many-body effects in this
exciting system.

This research was supported by ONR and NSF.  S.~T. Thompson
acknowledges the support of the ARO-MURI Fellowship Program.

$^*$Present address: Laboratoire Kastler Brossel, ENS, 24 rue
Lhomond, 75005 Paris, France.

\end{document}